%plain tex doc
\input phyzzx
\input epsf

\def\cOi{{\cal O}^\infty}
\def\tM{{\tilde M}}
\def\tP{{\tilde P}}

\def\oP{{\overline P}}
\def\nr{{n_r}}
\def\lr{{l_r}}
\def\anr{{\alpha_{n_r}}}
\def\bnr{{\beta_{n_r}}}

\rightline{UATP-00/03}
\centerline{\fourteenrm The Quantum States and the Statistical Entropy} 
\centerline{\fourteenrm of the Charged Black Hole}
\vskip 0.25in
\centerline{{\caps Cenalo Vaz}\footnote{\dagger}{Email: cvaz@ualg.pt.}}
\centerline{\it UCEH, Universidade do Algarve}
\centerline{\it Campus de Gambelas}
\centerline{\it P-8000 Faro, Portugal.}
\vskip 0.1in
\centerline{{\caps Louis Witten}\footnote{\ddagger}{Email: witten@physics.uc.edu}}
\centerline{\it Department of Physics}
\centerline{\it University of Cincinnati}
\centerline{\it Cincinnati, OH 45221-0011}
\vskip 0.25in
\centerline{\caps Abstract}
\vskip 0.1in
\noindent We quantize the Reissner-Nordstr\"om black hole using an adaptation of 
Kucha\v r's canonical decomposition of the Kruskal extension of the Schwarzschild 
black hole. The Wheeler-DeWitt equation turns into a functional Schroedinger equation 
in Gaussian time by coupling the gravitational field to a reference fluid or dust. The 
physical phase space of the theory is spanned by the mass, $M$, the charge, $Q$, 
the physical radius, $R$, the dust proper time, $\tau$, and their canonical momenta. 
The exact solutions of the functional Schroedinger equation imply that the difference 
in the areas of the outer and inner horizons is quantized in integer units. This 
agrees in spirit, but not precisely, with Bekenstein's proposal on the discrete horizon 
area spectrum of black holes. We also compute the entropy in the microcanonical 
ensemble and show that the entropy of the Reissner-Nordstr\"om black hole is 
proportional to this quantized difference in horizon areas.
%\vskip 0.2in

\noindent{\caps PACS:} 04.60.Ds, 04.70.Dy 
\vfil\eject

\noindent{\bf I. Introduction.}

Although the temperature of a black hole is exactly zero degrees Kelvin in classical general 
relativity, Bekenstein, in his 1972 thesis,${}^{1}$ proposed that black holes have a 
temperature and entropy, and should be treated as thermodynamic systems. The temperature 
and entropy of the black hole are known from semi-classical arguments${}^{2,3,4}$ to be 
fundamentally quantum mechanical in nature. Therefore, understanding their origins from a 
bona fide microcanonical ensemble of quantum states has come to be recognized as a 
challenge of considerable importance for an eventual theory of quantum gravity.

The earliest attempt at a microscopic theory of black holes was also due to Bekenstein. 
The argument roughly goes as follows.${}^{5,6}$ Using the Christodoulou-Ruffini${}^{7}$ process, 
it becomes clear that the horizon area operator of the black hole (in the case of multiple 
horizons, the area of the outer horizon) must be treated as an adiabatic invariant. Then, 
invoking the semi-classical Bohr-Sommerfeld quantization rules, Bekenstein concluded that 
the horizon area operator admits a discrete, equally spaced spectrum, ${\cal A}_n \approx nl_p^2$, 
where $l_p$ is the Planck length, and proceeded to use this spectrum as the rationale 
for dividing the horizon into cells of unit area which get added one by one and which have 
the same, small number of states, say $k$. The result is an estimate of the density of 
microstates, $\Omega \approx k^n$, or the entropy of the black hole, $S = \ln \Omega 
\approx n \ln k \approx \ln k ({\cal A}/l_p^2)$.  

This paper is a development of earlier work,${}^{8,9}$ where the quantum states and the total
entropy of the Schwarzschild black hole were recovered by combining the canonical reduction of 
spherical geometries by Kucha\v r${}^{10}$ with the coupling to an external reference fluid 
(or dust), originally proposed by Kucha\v r and Torre${}^{11}$ and Kucha\v r and Brown.${}^{12}$
The functional Schroedinger equation (in the dust proper time) that was obtained by this procedure 
described the more general problem of inhomogeneous dust collapse. It was simplified by holding 
the mass of the hole constant, independent of the spatial coordinate, and could then 
be easily solved throughout the Kruskal manifold. This gave precisely Bekenstein's area 
quantization law. The coupling to dust may be thought of either as a way to impose coordinate 
conditions (ref. [11]), which is the point of view we take here, or as a realistic material 
medium (ref. [12]).

Our goal in this paper is to extend this analysis to the charged, Reissner-Nordstr\"om black 
hole. We will see that, as in the Schwarzschild case, the quantization leads to a derivation of
the statistical properties of the black hole. In particular, the entropy will turn out to be 
proportional to the difference between the outer and inner horizon areas and will be quantized 
in integer units. Thus, although we do not recover Bekenstein's area quantization, our result 
is in keeping with its spirit in as much as it is commensurate with an ``area quantization'' 
law. As the charge, $Q$, approaches zero the results will approach those obtained for the 
uncharged Schwarzschild black hole and in the limiting case, as the black hole becomes extremal, 
the entropy will approach zero. 

The Reissner-Nordstr\"om solution is given in curvature coordinates by the line element 
$$ds^2~~ =~~ F(R) dT^2~ -~ F^{-1}(R) dR^2~ -~ R^2 d\Omega^2,\eqno(1.1)$$
where $d\Omega$ is the ordinary unit sphere, $R$ is the physical radius, the coefficient
$F(R)$ has the form
$$F(R)~~ =~~ 1~ -~ {{2M} \over R}~ +~ {{Q^2} \over {R^2}},\eqno(1.2)$$
and the electromagnetic potential is
$$A~~ =~~ {Q \over R} dT.\eqno(1.3)$$
The vector field $\partial/\partial T$ is a Killing vector field of the metric. It is 
time-like in the interior (regions IV and V, see the Penrose diagram in figure 1) and in the 
exterior (regions I and II), but space-like in region III. 

An important feature of the maximal extension of this geometry is that the inner horizon is 
a Cauchy horizon for spatial sections, $\Sigma$ (see figure 1).
\vskip 0.2in
\centerline{\epsfysize 2.5in \epsfbox{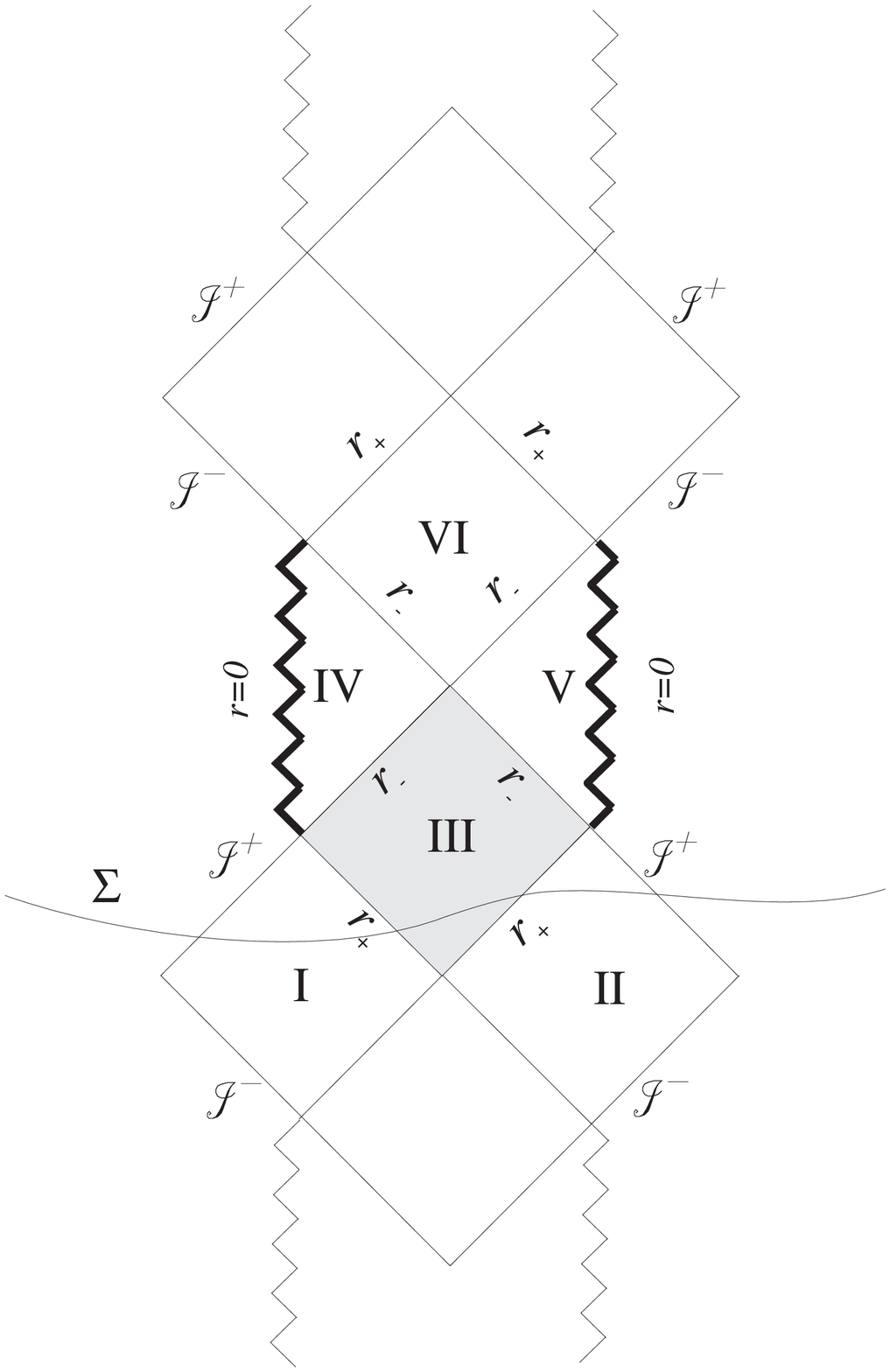}}
\centerline{\Tenpoint {\bf Fig. 1:} The extended Reissner-Nordstr\"om geometry}
\vskip 0.1in
\noindent What this means is that if data were given on an initial 
hypersurface, $\Sigma$, the Cauchy development will be able to predict only what occurs in 
regions I, II and III and not beyond the inner horizons, in regions IV, V of the diagram. 
Any event in regions IV and V of the space-time is influenced not simply by the given data 
and evolution but by additional data on the singularities themselves, which are impossible 
to control. Such a situation does not arise for the black hole (see figure 2) where spatial 
sections are able to cover all of space-time until the singularity is reached and the 
Cauchy development is able to predict what occurs everywhere, once data is given on an 
initial hypersurface.
\vskip 0.2in
\centerline{\epsfysize 1.5in \epsfbox{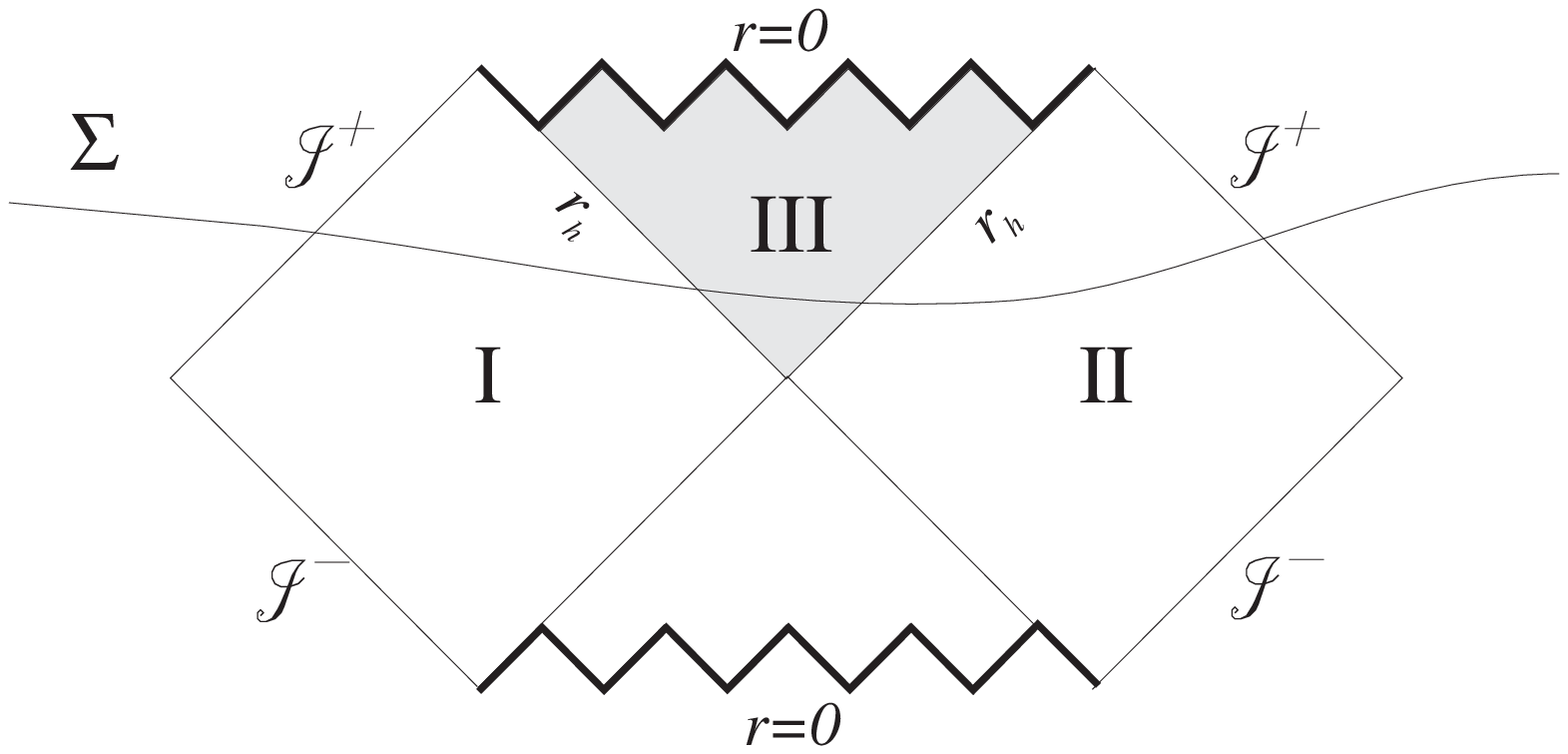}}
\centerline{\Tenpoint {\bf Fig. 2:} The extended Schwarzschild geometry}
\vskip 0.1in
\noindent The non-existence of well-defined Cauchy surfaces in regions IV and V of the 
extended space-time means that the canonical theory is impossible to define there. We will 
therefore not attempt to identify the quantum theory in regions IV and V of figure 1. Instead
we will ask what are the consequences of spatial diffeomorphism invariance in the interior. 
We will see that the wave-functional is required to be either vanishing or constant there. 

We will consider the Einstein-Maxwell-dust system described by the action
$$S~~ =~~ -~ { 1 \over {16\pi}} \int d^4 x \sqrt{-g} [{\cal R} - F_{\mu\nu} F^{\mu\nu}]~ 
-~ {1 \over {8\pi}} \int d^4 x \sqrt{-g} \epsilon(x) \left[g_{\alpha\beta} U^\alpha U^\beta + 1
\right] \eqno(1.4)$$
in the general spherically symmetric space-time
$$ds^2~~ =~~ N^2 dt^2~ -~ L^2 (dr - N^r dt)^2~ -~ R^2 d\Omega^2,\eqno(1.5)$$
where $N(t,r)$ and $N^r(t,r)$ are respectively the lapse and shift functions, $R(t,r)$ 
is the physical radius or curvature coordinate, $\epsilon(t,r)$ is the density of 
the collapsing dust in its proper frame, ${\cal R}$ is the scalar curvature and 
$U^\alpha$ are the components of the dust proper velocity.

This paper is organized as follows. In section II we summarize the general canonical 
formalism for spherically symmetric space-times, stating the canonical form of the 
action, the appropriate fall-off conditions to be imposed on the canonical variables at 
infinity and the boundary terms. In section III we recast the action in terms of a new 
chart composed of the mass, the curvature coordinate, the dust proper time and their 
conjugate momenta. Sections II and III will closely follow Kucha\v r's original 
reasoning,${}^{10}$ which will be adapted to suit the charged black hole geometry. In 
section IV we obtain and solve the Wheeler-DeWitt equation, subject to spatial 
diffeomorphism invariance,  for the Reissner-Nordstr\"om black hole, thus recovering 
the spectrum of the black hole. In section V we recast our solution in a suitable form 
and obtain the total entropy and the Hawking temperature in the microcanonical ensemble. 
We conclude in section VI with a few comments on the results obtained and the assumptions 
that went into their making.
\vskip 0.2in

\noindent{\bf II. Hamiltonian Reduction.}

The gravitational part of the action in (1.4) was recast by Kucha\v r${}^{10}$ into 
the form 
$$S^g~~ =~~ \int dt \int dr \left[P_L {\dot L}~ +~ P_R {\dot R}~ -~ N H^g - N^r H^g_r \right]~~ +~~ 
S^g_{\partial \Sigma},\eqno(2.1)$$
where 
$$\eqalign{P_L~~ &=~~ {R \over N} \left[ - {\dot R} + N^r R'\right]\cr P_R~~ &=~~ 
{1 \over N} \left[-L{\dot R} - {\dot L} R + (N^r L R)'\right]\cr}\eqno(2.2)$$
are the momenta conjugate to $L$ and $R$ respectively, and 
$$\eqalign{H^g~~ &=~~ -\left[{{P_L P_R} \over {R}} - {{LP_L^2} \over {2R^2}}\right]
+ \left[-{L\over 2} - {{R^{'2}}\over {2L}} + \left({{RR'} \over L}\right)'\right]\cr 
H^g_r~~ &=~~ R'P_R - L P_L'.\cr}\eqno(2.3)$$
Again, $S^g_{\partial\Sigma}$ is a surface term that is required to cancel unwanted 
boundary terms in the variations of the canonical variables and must be determined 
after specifying reasonable fall-off conditions on the canonical variables and the 
Lagrange multipliers of the theory. Kucha\v r's fall-off conditions are well suited to 
the exterior of the maximally extended Reissner-Nordstr\"om geometry and we shall 
adopt them here. They read
$$\eqalign{L(t,r)~~ &=~~ 1~ +~ M_\pm(t) |r|^{-1}~ +~ \cOi(|r|^{-1-\epsilon})\cr R(t,r)~~ 
&=~~ |r|~ +~ \cOi(|r|^{-\epsilon})\cr P_L(t,r)~~ &=~~ \cOi(r^{-\epsilon})\cr P_R
(t,r)~~ &=~~ \cOi(|r|^{-1-\epsilon})\cr N(t,r)~~ &=~~ N_\pm(t)~ +~ \cOi(|
r|^{-\epsilon})\cr N^r(t,r)~~ &=~~ \cOi(|r|^{-\epsilon})\cr}\eqno(2.4)$$
and, with them, the boundary action required is easily seen to be
$$S^g_{\partial\Sigma}~~ =~~ - \int dt [N_+(t) M_+(t)~ +~ N_-(t) M_-(t)].\eqno(2.5)$$
Kucha\v r has emphasized that $N_\pm(t)$ must be considered as {\it prescribed}
functions of the label time coordinate, otherwise a variation of the total action would 
also lead to the conclusion that the energy of the system at infinity is exactly zero. The 
fact that the $N_\pm(t)$ are prescribed functions will be exploited below to set the 
parametrization clocks at infinity.

A straightforward canonical reduction of the electromagnetic term in (1.4) with 
the ansatz in (1.5) gives
$$S^{em}~~ =~~ \int dt \int dr \left[P_A {\dot A}_r~ -~ N H^{em} -  N^r H^{em}_r - 
\phi P_A'\right]~ +~  S^{em}_{\partial\Sigma},\eqno(2.6)$$
where 
$$P_A~~ =~~ {{R^2} \over {NL}} \left[{\dot A}_r~ -~ A_t'\right]\eqno(2.7)$$
is conjugate to $A_r$,
$$\eqalign{H^{em}~~ &=~~ +~ {{LP_A^2} \over {2R^2}}\cr H^{em}_r~~
&=~~ -~ A_r P_A',\cr}\eqno(2.8)$$
and we have defined 
$$\phi(t,r)~~ =~~ -~ A_t(t,r)~ +~ N^r(t,r) A_r(t,r).\eqno(2.9)$$
If we adopt the following fall-off conditions
$$\eqalign{A_r(t,r)~~ &=~~ \cOi(|r|^{-1-\epsilon})\cr P_A(t,r)~~ &=~~ Q_\pm(t)~ +~ 
\cOi(|r|^{-\epsilon})\cr \phi(t,r)~~ &=~~ \phi_\pm(t)~ +~ \cOi(|r|^{-\epsilon}),\cr}
\eqno(2.10)$$
then the electromagnetic surface term is of the form
$$S^{em}_{\partial \Sigma}~~ =~~ -\int dt~ [\phi_+ Q_+~ -~ \phi_- Q_-]. \eqno(2.11)$$
$Q_\pm$ will turn out to be the electric charge. Once again, to avoid a neutral solution 
($Q=0$) we must treat $\phi_\pm(t)$ as prescribed functions of the label time coordinate. 
A gauge choice that is consistent with the Reissner-Nordstr\"om solution is $\phi_\pm(t) = 0$
and so this term vanishes.

Let us now consider the dust action in (1.4). Dust is described by eight space-time 
scalars, $\epsilon$, $\tau$, $Z^k$ and $W_k$ ($k~ \in~ \{1,2,3\}$). The physical 
interpretation of these variables which follows from an analysis of the equations of motion 
were given in ref. [12] and will be summarized here for completeness. $\tau$ is the 
proper time measured along particle flow lines, $Z^k$ are the comoving coordinates of 
the dust, $W^k$ are the spatial components of the four velocity in the dust frame, and 
$\epsilon$ is the dust proper energy density. All these scalars are assumed to be 
functions of the space-time coordinates. In particular, the four variables, $Z^K = 
(\tau, Z^k)$, are independent functions, ${\rm det}|Z^K_{~~,\mu}| \neq 0$, and the 
four-velocity of the dust particles may be defined by its decomposition in the cobasis 
$Z^K_{~~,\mu}$ by
$$U_\mu~~ =~~ -\tau_{,\mu}~ +~ W_k Z^k_{~~, \mu}.\eqno(2.12)$$
In the spherically symmetric geometry described by (1.5), the dust action may be 
cast into the form
$$S^d~~ =~~ \int dt \int dr  \left[P_\tau {\dot \tau} + P_k {\dot Z}^k - N H^d - N^r H^d_r 
\right],\eqno(2.13)$$
where
$$\eqalign{P_\tau~~ &=~~ {{LR^2} \over N} \epsilon(r,t) \left[-U_t + N^r U_r\right]\cr 
P_k~~ &=~~  - W_k P_\tau\cr}\eqno(2.14)$$
are the momenta conjugate to the the dust proper time and the frame variables respectively, 
and
$$\eqalign{H^d~~ &=~~ P_\tau \sqrt{1 + {{U_r^2} \over {L^2}}}\cr H^d_r~~ &=~~ 
- U_r P_\tau.\cr}\eqno(2.15)$$
The expression for $H^d$ in (2.15) is obtained upon exploiting the fact that $\epsilon(t,r)$ 
is a Lagrange multiplier and therefore $\delta {\cal L} / \delta \epsilon = 0$.

Putting the three components of our system together, we have the Hamiltonian form of the 
total action in (1.4) with the ansatz in (1.5). It reads,
$$S~~ =~~ \int dt \int dr [P_L {\dot L} + P_R {\dot R} +~ P_A {\dot A}_r + P_\tau {\dot \tau} 
+ P_k {\dot Z}^k - N H - N^r H_r - \phi G]~ +~ S_{\partial\Sigma},\eqno(2.16)$$
where the boundary action, $S_{\partial\Sigma}=S^g_{\partial\Sigma}$ and is given by the 
right hand side of (2.5). The full super-Hamiltonian and supermomentum constraints are 
given by
$$\eqalign{H~~ &=~~ -~ \left[{{P_L P_R} \over {R}} - {{LP_L^2} \over {2R^2}} \right] + 
\left[-{L\over 2} - {{R^{'2}}\over {2L}} + \left({{RR'} \over L} \right)'\right]\cr 
&~~~~~~ +~ {{L P_A^2} \over {2R^2}}~ +~ P_\tau \sqrt{1 + {{U_r^2} \over {L^2}}}~~ 
\approx~~ 0,\cr H_r~~ &=~~  R' P_R - L P_L' - A_r P_A' - U_r P_\tau~~ \approx~~ 0\cr}
\eqno(2.17)$$
and the electromagnetic constraint by,
$$G~~ =~~ P_A'~~ \approx~~ 0.\eqno(2.18)$$
The constraints may be further simplified by requiring that the dust be non-rotating 
and that its motion be described with respect to the frame orthogonal foliation. Then 
we may impose the additional constraints${}^{12}$ $P_k =0$. When they are applied  
as restrictions on the state functional $\Psi[\tau,Z,g,A]$, they imply that the state 
$\Psi$ does not depend on the frame variables $Z^k$. The Hilbert space is then composed
of  state functionals, $\Psi[\tau,g,A]$, and the quantum theory is described by imposing 
the classical constraints as operator conditions on them,
$${\hat H}(\tau,g,A) \Psi[\tau,g,A]~~ =~~ 0~~ =~~ {\hat H}_r(\tau,g,A) \Psi[\tau,g,A],
\eqno(2.19)$$
after a suitable operator ordering has been found. Using (2.12) with $W_k=0$, the 
classical constraints take the form
$$\eqalign{H~~ &=~~ -~ \left[{{P_L P_R} \over {R}} - {{LP_L^2} \over {2R^2}} \right] + 
\left[-{L\over 2} - {{R^{'2}}\over {2L}} + \left({{RR'} \over L} \right)'\right]\cr 
&~~~~~~ +~ {{L P_A^2} \over {2R^2}}~ +~ P_\tau \sqrt{1 + {{\tau^{'2}} \over {L^2}}}~~ 
\approx~~ 0,\cr H_r~~ &=~~  R' P_R - L P_L' - A_r P_A' + \tau' P_\tau~~ \approx~~ 0\cr
G~~ &=~~ P_A'~~ \approx~~ 0.\cr}\eqno(2.20)$$
Not only are they not decoupled, making them difficult to solve, but the phase space 
variables are not transparent and ``natural'' to the black hole problem.
The spatial hypersurfaces must eventually be embedded in the metric given by (1.1) and 
(1.2) and this line element is completely characterized by the mass, $M$, and the charge,
$Q$, of the black hole. Kucha\v r${}^{10}$ has shown both how these quantities are determined
by the canonical data as well as  how the hypersurface embedding in the space-time 
may be deduced from the values of the phase space coordinates at any point. This leads 
to a reformulation of the constraints (2.20) in terms of more transparent variables,
the mass, the charge, the physical radius, the dust proper time and their conjugate 
momenta which we describe in the following section.
\vskip 0.2in

\noindent{\bf III. New Variables and New Constraints.}

Substituting the foliation, $T=T(t,r)$, $R=R(t,r)$, into the line element given 
in (1.1) and comparing it with the ADM form of the same, {\it i.e.,} (1.5), we find
$$\eqalign{L^2~~ &=~~ -FT^{'2}~ +~ F^{-1}R^{'2}\cr L^2 N^r~~ &=~~ -F T'{\dot T}
~ +~ F^{-1} R'{\dot R}\cr N^2 - L^2 N^{r2}~~ &=~~ F{\dot T}^2~ -~ F^{-1} {\dot R}^2.
\cr}\eqno(3.1)$$
These identities may be easily solved for the lapse and shift functions, giving
$$\eqalign{N~~ &=~~ {{R'{\dot T} - T'{\dot R}} \over {\sqrt{-F T^{'2} + 
F^{-1}R^{'2}}}}\cr N^r~~ &=~~ {{-FT'{\dot T} + F^{-1} R'{\dot R}} \over {-F T^{'2}
+ F^{-1} R^{'2}}}.\cr}\eqno(3.2)$$ 
Defined with the positive square-root, the lapse function is positive in all three 
regions of interest with label time going to the future. Substituting the expressions
(3.2) into (2.2), one finds
$$T'~~ =~~ -~ {{LP_L} \over {RF}},\eqno(3.3)$$
which can be inserted into the expression for $L^2$ in (3.1) to give 
$$F(R)~~ =~~ {{R^{'2}} \over {L^2}}~ -~ {{P_L^2} \over {R^2}}.\eqno(3.4)$$
Let us, for the present, work with the mass function $\tM$, which we define by 
$F(R) = 1-2\tM/R$. Comparing this with (1.2) it is clear that $\tM$ must be related 
to the mass and charge of the black hole by
$$\tM~~ =~~ M~ -~ {{Q^2} \over {2R}},\eqno(3.5)$$
if we are to recover the Reissner-Nordstr\"om black hole. $\tM$ as determined by 
the canonical data is
$$\tM~~ =~~ {R \over 2}\left[1 - {{R^{'2}} \over {L^2}} + {{P_L^2} \over 
{R^2}} \right].\eqno(3.6)$$
Note that $\tM(r)$ is a local function of the canonical data as is $T'(r)$. If we now 
proceed to compute the Poisson brackets between $\tM(r)$ and $T'(r)$ from the 
fundamental Poisson brackets implied by the Liouville form in (2.16) we will see that 
$-T'(r)$ can be interpreted as the momentum conjugate to $\tM(r)$. Henceforth we 
will refer to it as $\tP_M(r)$, {\it i.e.,}
$$\tP_M~~ =~~ -T'~~ =~~ {{LP_L} \over {RF}}.\eqno(3.7)$$
Now, while the pair $\{\tM,\tP_M\}$ has vanishing Poisson brackets with $R$, it 
does not have vanishing Poisson brackets with $P_R$. We would like to find a 
transformation that takes the chart $\{R, P_R, L, P_L, A_r, P_A, \tau, P_\tau\}$ 
to a new chart explicitly involving the mass and charge of the system, $\{R, 
\tP_R, \tM, \tP_M, Q, P_Q, \tau, P_\tau\}$, subject to (3.6), (3.7) and 
$Q(r) = P_A(r)$, $A_r(r) = - P_Q(r)$. The last two relations constitute an 
elementary exchange of coordinates and momenta. The four conditions provide sufficient 
information to obtain a generating functional for the canonical transformation, 
which can be given in terms of the original phase space coordinates as
$${\cal F}[R,P_R,L,P_L,A_r,P_A]~~ =~~ \int dr \left[ A_r P_A + L P_L + {{RR'} \over 2} 
\ln|{{RR' - LP_L} \over {RR' + LP_L}}| \right],\eqno(3.8)$$
With the help of (3.6) and (3.7) we compute $\tP_R(r)$ directly from (3.8); it is
$$\tP_R(r)~~ =~~ P_R - {{LP_L} \over {2R}} - {{LP_L}\over {2RF}} - {1 \over {RL^2 F}}
[(RR')(LP_L)' - (RR')'(LP_L)].\eqno(3.9)$$
The fall-off conditions can be applied to show that the generating functional 
in (3.8) is well defined near infinity. It can also be shown to stay finite at the 
horizons. On the other hand, the transformation from the old chart to the new is 
invertible everywhere except at the horizons.

Thus, following Kucha\v r's reasoning for the Schwarzschild black 
hole${}^{10}$, we have introduced the mass and charge as dynamical 
variables on the phase space. We shall now re-express our constraints in (2.20) in 
terms of the new chart. Again, Kucha\v r has pointed the way: it follows from 
expression (3.6) for $\tM(r)$ and expressions (2.3) that 
$$\tM'~~ =~~ -~ {{R'} \over {L}} H^g~ -~ {{P_L} \over {RL}} H^g_r.\eqno(3.10)$$
This allows us to write the gravitational part of the super-Hamiltonian 
and supermomentum constraints in terms of the new variables. Using the transformations,
it is easy to see that (2.3) becomes 
$$\eqalign{H^g_r~~ &=~~ R'\tP_R~ +~ \tM' \tP_M\cr H^g~~ &=~~ -\left[{{\tM'F^{-1} R' + 
F\tP_M \tP_R} \over L} \right]\cr},\eqno(3.11)$$
where, expressed in terms of the new variables,
$$\eqalign{F~~ &=~~ 1~ -~ {{2\tM} \over R}\cr L^2~~ &=~~ - F\tP_M^2~ +~ F^{-1} R^{'2}.
\cr} \eqno(3.12)$$
Furthermore, the constraints $H \approx 0$, $H_r \approx 0$ and $G \approx 0$ in (2.20) 
together imply that 
$$\tM'~~  \approx~~ {{Q^2 R'} \over {2R^2}} + P_\tau {{R'} \over {L^2}}\sqrt{L^2 
+ \tau^{'2}} + {{F\tau' \tP_M P_\tau} \over {L^2}},\eqno(3.13)$$
showing that 
$$\tM~~ \approx~~ M~ -~ {{Q^2} \over {2R}},\eqno(3.14)$$
where $M'$ is given by the last two terms on the right hand side of (3.13). This 
compares with equation (3.5) and shows that to recover the Reissner-Nordstr\"om 
black hole we must further impose homogeneity via the constraint $M'(r) = 0$. It 
is not surprising that this condition has to be enforced by hand and does not 
follow directly from the constraints. In introducing non-rotating dust, we have 
introduced an extra degree of freedom in the theory. Thus the problem, as it has 
been set up here, actually describes the more general problem of gravitational 
collapse of inhomogeneous dust before the constraint $M'(r)=0$ is 
imposed. The black hole is treated as a special case of the general problem.

Let us now turn to the boundary term on the right hand side of (2.5). As mentioned, 
$N_\pm(t)$ must be treated as prescribed functions of $t$. The freedom in choosing 
this function can be combined with the freedom we have of setting the dust proper 
time at infinity to correspond to the parametrization clocks there. The lapse function 
is the rate of change of the proper time with the coordinate time at infinity, so 
we set $N_\pm(t) = \pm {\dot \tau}_\pm(t)$ to write
$$S_{\partial\Sigma}~~ =~~ - \int dt~ [M_+ {\dot \tau}_+~ -~ M_- {\dot \tau}_-].
\eqno(3.15)$$
It is linear in the time derivatives, ${\dot \tau}_\pm$, and defines a one form,
$$\eqalign{-~ [M_+ \delta \tau_+~ -~ M_- \delta \tau_-]~~ &=~~ - \int dr (\tM \delta 
\tau)'\cr &=~~  - \int dr [\tM' \delta \tau - \tau' \delta \tM + \delta(\tM\tau')].
\cr}\eqno(3.16)$$
The first two terms on the right hand side may be absorbed into the Liouville form
of the hypersurface action (they modify the canonical momenta) and the last term is 
an exact form which can be dropped. The action is expressed entirely as
a hypersurface action. Defining $\oP_\tau = P_\tau - \tM'$ and $\oP_M = \tP_M + \tau'$, 
we may write it as 
$$S~~ =~~ \int dt \int dr [\tP_R {\dot R} + \oP_M {\dot \tM} +~ P_Q {\dot Q} + \oP_\tau 
{\dot \tau}  - N H - N^r H_r - \phi G],\eqno(3.17)$$
where the constraints in the new chart reads
$$\eqalign{H~~ &=~~ -\left[{{\tM' F^{-1} R'~ +~ F(\oP_M-\tau') \tP_R} \over L}\right]~ 
+~ {{LQ^2} \over {2R^2}}\cr &~~~~~~~~~~ +~ (\oP_\tau + \tM') \sqrt{1 + {{\tau^{'2}}\over 
{L^2}}}~~ \approx~~ 0\cr H_r~~ &=~~ R'\tP_R~ +~ \tM'\oP_M~ +~ Q'P_Q~ +~ \tau' \oP_\tau~~ 
\approx~~ 0\cr G~~ &=~~ Q'~~ \approx~~ 0,\cr}\eqno(3.18)$$
and where $L^2$ is given in (3.12). A final point transformation,
$$\eqalign{R~~ &=~~ R,~~~~~~~~~~ M~~ =~~ \tM~ +~ {{Q^2} \over {2R}}\cr \oP_M~~ &=~~ 
\oP_M ,~~~~~~~~~~ \oP_R~~ =~~ \tP_R~ +~ {{Q^2\oP_M} \over {2R^2}},\cr}\eqno(3.19)$$
suggested by (3.14), would express the action above in terms of the true mass, $M$.
Furthermore, the Poisson brackets of the new phase space coordinates with the 
constraints would yield the canonical equations of motion in terms of them. 

If the supermomentum constraint is now used to eliminate $\oP_M$ in the super-Hamiltonian,
the latter constraint takes a relatively simple form
$$(\oP_\tau + \tM')^2 ~ +~ F\left[\oP_R - {{Q^2\tau'} \over {2R^2}}\right]^2~ -~ 
{{M^{'2}} \over F}~~ \approx~~ 0,\eqno(3.20)$$
and specializing to the Reissner-Nordstr\"om black hole by allowing only the 
homogeneous mode to survive, $M'(r) = 0$, gives the 
final form 
$$\left[ \oP_\tau + {{Q^2 R'} \over {2R^2}}\right]^2~ +~ F\left[\oP_R - {{Q^2\tau'} 
\over {2R^2}}\right]^2~~ \approx~~ 0.\eqno(3.21)$$
It remains to impose the constraint as an operator equation on the state functional 
$\Psi[\tau,R,Q]$ and solve the resulting functional Schroedinger equation. The solutions
must also obey spatial diffeomorphism invariance and this is imposed by the 
second constraint in (3.18). Acceptable solutions are the topic of the next 
section.
\vskip 0.2in

\noindent{\bf IV. Quantization.}

The quantum state of the black hole on a hypersurface $\tau(r)$, $R(r)$ at a label time 
$t$ is described by a state functional $\Psi[\tau,R,Q]$ over the configuration space, 
with the canonical momenta acting upon it as functional differential operators. The 
configuration space can be seen to admit the inverse metric $\gamma^{ab} = {\rm diag}
(1,F)$, $\gamma_{ab} = {\rm diag} (1,1/F)$. The metric is positive definite in the external
(and internal) regions, and indefinite in the dynamical region. Furthermore, it is a 
flat metric and can be brought to manifestly flat form by a coordinate transformation. 
In the external region, this transformation takes the form 
$$\eqalign{R_*~~ &=~~ \int {{dR} \over {\sqrt{F(R)}}}~~ =~~ \int {{RdR} \over 
{\sqrt{R^2 - 2MR + Q^2}}}\cr &=~~ R\sqrt{F(R)} + M\ln |R-M+R\sqrt{F(R)}|\cr}\eqno(4.1)$$
and in the dynamical region it is
$$\eqalign{R_*~~ &=~~ \int {{dR} \over {\sqrt{-F(R)}}}~~ =~~ \int {{RdR}\over 
{\sqrt{2MR - R^2 - Q^2}}}\cr &=~~ -R\sqrt{-F(R)} - M \sin^{-1}\left[{{M-R}\over 
{\sqrt{M^2-Q^2}}}\right].\cr}\eqno(4.2)$$
These definitions may be modified by additive constants so as to turn $R_*$ into a continuous
variable. The classical constraints can now be expressed in terms of the momentum conjugate to 
$R_*$ in each region. They are
$$\eqalign{R_*'\oP_*~ +~ M'\oP_M~ +~ Q'P_Q~ +~ \tau' \oP_\tau~~ &\approx~~ 
0\cr \left[ \oP_\tau + {{Q^2 R'} \over {2R^2}}\right]^2 ~ \pm~ \left[\oP_* - {{Q^2\sqrt{
\pm F(R)}\tau'} \over {2R^2}}\right]^2~~ &\approx~~ 0,\cr}\eqno(4.3)$$
where the $\pm$ signs refer to the exterior and the dynamical region respectively, and 
must be imposed as operator constraints on the wave functional. In the new, 
manifestly flat, configuration space, we represent the momenta by the functional 
derivatives
$$\eqalign{{\hat \oP}_\tau(r) ~~ &=~~ -i {\delta \over {\delta \tau(r)}}\cr 
{\hat \oP}_*(r) ~~ &=~~ -i {\delta \over {\delta R_*(r)}}\cr}\eqno(4.4)$$
and consider the solutions of the equation
$$\left\{\left[ \delta_\tau + i{{Q^2 R'} \over {2R^2}}\right]^2 ~ \pm~ \left[\delta_* 
- i{{Q^2\sqrt{\pm F(R)}\tau'} \over {2R^2}}\right]^2\right\}\Psi[\tau,R_*,Q]~~ =~~ 0
\eqno(4.5)$$
that respect the diffeomorphism constraint, 
$$\left\{R_*'\delta_*~ +~ \tau' \delta_\tau\right\} \Psi[\tau,R_*,Q]~~ =~~ 0, \eqno(4.6)$$
where we have already imposed the two requirements, ${\hat Q}' \Psi = 0$ and ${\hat M}' 
\Psi = 0$, the first of which is simply the electromagnetic constraint and the need for 
the second arises, as we have mentioned earlier, from the fact that we have added a degree of 
freedom to the system in the form of dust, which must be constrained in order to 
describe the Reissner-Nordstr\"om black hole. Equations (4.5) and (4.6) define the quantum 
theory whose Hilbert space is ${\cal H}:= {\cal L}^2({\rm {\bf R}}, dR_*)$ with inner 
products 
$$\langle \Psi_1, \Psi_2\rangle~~ =~~ \int_{M \ln \sqrt{M^2 - Q^2}}^\infty dR_* \Psi_1^\dagger
\Psi_2\eqno(4.7)$$
in the exterior region and 
$$\langle \Psi_1, \Psi_2\rangle~~ =~~ \int_{-{{\pi M} \over 2}}^{{\pi M} \over 2} dR_* 
\Psi_1^\dagger \Psi_2\eqno(4.8)$$
in the dynamical region.

Now it is readily verified, by taking functional derivatives, that any solution of 
equation (4.5) can be written in the form
$$\Psi[\tau,R_*,Q]~~ =~~ \exp\left[i\int dr\left({{Q^2} \over {2R(r)}}\right)'\tau(r)\right]
\Psi_o[\tau,R_*],\eqno(4.9)$$
where $\Psi_o$ is a solution of the {\it free} equation
$$(\delta_\tau^2 \pm \delta_*^2)\Psi_o[\tau,R_*]~~ =~~ 0.\eqno(4.10)$$
We have, once again, used the $\pm$ signs to refer to the exterior and the dynamical regions 
respectively. Not surprisingly, (4.10) was obtained for the Schwarzschild black hole 
in refs. [8,9]. Two features of this equation are worthy of mention. For one, a consequence 
of the signature change in the configuration space metric as we move from the external region 
to the dynamical region is that the ``equation of motion'' goes from being elliptic to 
hyperbolic. For another, (4.10) is decoupled implying that it may be solved independently at 
each point, labeled by $r$, of the spatial hypersurface. In the exterior, then, the 
solutions will be exponentially decaying, but they will oscillate in the dynamical 
region and in each region they will have the form
$$\Psi_o[\tau,R_*]~~ =~~ \prod_r \Psi_o[\tau(r),R_*(r)].\eqno(4.11)$$
Thus we may write a general solution of (4.5) as
$$\Psi[\tau,R_*,Q]~~ =~~ \exp\left[i\int dr\left({{Q^2} \over {2R(r)}}\right)'\tau(r)
\right] \prod_r \Psi_o[\tau(r),R_*(r)].\eqno(4.12)$$
Again, the solution will obey spatial diffeomorphism invariance if and only if 
$\Psi_o[\tau,R_*]$ obeys (4.6).

Consider the solutions in the exterior ($F>0$). The general positive energy solution that 
is well behaved in the entire range of $R_*$ is given by
$$\Psi_o[\tau(r),R_*(r)]~~ =~~ c(M,Q) \exp\left[-i E (\tau(r)- i R_*(r))\right],
\eqno(4.13)$$
where $c(M,Q)$ is a mass and charge dependent constant. However, upon considering the 
action of the spatial diffeomorphism invariance constraint on $\Psi_o[\tau,R_*]$ defined in 
(4.11), we conclude that $E(\tau' - i R_*')\Psi_o = 0$. For positive energy solutions, this 
condition is met by $\tau' = 0 = R_*'$, by $E=0$ or by $c(M,Q) = 0$. However, $R_*' = 0$ 
implies from (4.1) that $R'=0$ and this is unacceptable in the exterior because $L^2$ given 
in (3.12) is required to be positive definite. We could take $E = 0$ and $\Psi_o$ would be 
a constant, $c(M,Q)$, but this solution would not be ${\cal L}^2$ according to (4.7).
We conclude that $c(M,Q)$ must vanish and consequently that the wave functional, 
$\Psi_o[\tau,R_*]$, is identically zero in the exterior of the black hole. 

In the dynamical region, $F<0$, the situation is completely different. The ``equation of 
motion'' is hyperbolic and the solutions are oscillatory. The general positive energy solution 
is now given by 
$$\Psi_o[\tau(r),R_*(r)]~~ =~~ a(M,Q) e^{-iE[\tau(r)+R_*(r)]} + b(M,Q) e^{-iE[\tau(r)-R_*(r)]},
\eqno(4.14)$$
where $a(M,Q)$ and $b(M,Q)$ are mass and charge dependent constants. Because the 
wave functional in the exterior vanishes, $\Psi_o[\tau(r),R_*(r)]$ in (4.14) must vanish 
on the outer horizon, $R_* = \pi M/2$. This gives
$$b(M,Q)~~ =~~ - a(M,Q) e^{-i\pi E M}\eqno(4.15)$$
and
$$\Psi_o[\tau(r),R_*(r)]~~ =~~ a(M,Q) \left[e^{-iE[\tau(r)+R_*(r)]}~ -~ e^{-i\pi E M}
e^{-iE[\tau(r)-R_*(r)]}\right].\eqno(4.16)$$
Further boundary conditions come from matching the wave-functional, $\Psi[\tau,R_*]$ above, 
with its counterpart in the interior on the inner horizon. As we have mentioned in the 
introduction, the inner horizon is a Cauchy horizon for spatial sections. It is impossible 
as far as we know to define the dynamics consistently here because the Cauchy data must be 
supplemented by boundary conditions on the singularities which are impossible to give. We 
would expect that, since there is no dynamics within the inner horizon and since the 
diffeomorphism constraint must hold there, the wave function will be constant or equal to 
zero everywhere. If we take it to be identically zero and require that $\Psi_o[\tau,R_*]$ 
vanish on the inner horizon as well, one sees that the energy is quantized according to 
the simple relation $EM = n$, where $n$ is a non-negative integer. 

Yet, a vanishing wave-functional in the interior may not be the only possibility. Even though 
we can make no statements about the evolution equations, it is reasonable to expect that 
spatial diffeomorphism invariance of the wave-functional in the interior is respected according 
to (4.6). In the privileged foliation defined by the dust proper time, $\tau' = 0$ but $R' 
\neq 0$. In fact, $R'$ can never be zero in the static regions. It follows from (4.6) that 
the wave-functional on constant proper time hypersurfaces must be constant, {\it i.e.,} 
it may depend only on $\tau$. One easily verifies, for example, that the charge dependent 
phase of (4.9) vanishes on every leaf of this foliation. In order for the wave-functional 
in the interior to be consistently matched with the wave-functional in the dynamical 
region it must have the form 
$$\Psi^{int}[\tau,R_*]~~  =~~ \prod_r d(M,Q) e^{-iE\tau}.\eqno(4.17)$$
This, of course, does not respect the super-Hamiltonian constraint in (4.5), but 
we do not wish to impose it in the interior. Matching the wave-functionals in the 
interior and in the dynamical region one finds that the energy is quantized in half 
integer units of the Planck mass (as opposed to integer units when $d(M,Q)=0$) according
to $EM=(n+1/2)$. The ground state energy is non-zero excluding, as we will see, the
extremal solution as well as the Minkowski vacuum. Moreover, it does not reduce to 
the spectrum of the Schwarzschild black hole${}^{8,9}$ in the $Q\rightarrow 0$ limit. It 
will not change our counting of the black hole states and we will not pursue this 
case further in this paper.

We must now address the physical meaning of the momentum conjugate to the dust proper time in 
region III, {\it i.e.,} what is the functional dependence of the proper ``energy'', $E$, on 
the mass and charge of the hole? By our coupling to dust, the time coordinate is always chosen 
to coincide with the proper time, $\tau$, of a freely falling observer. It can be expressed 
in terms of the curvature coordinates as
$$\tau(T,R)~~ =~~ T~ +~ \int dR {{\sqrt{1-F(R)}}\over F(R)}\eqno(4.18)$$
in the (static) regions I, II, IV and V, and 
$$\tau(R)~~ =~~ R_*~~ =~~ \int {{dR} \over {\sqrt{-F(R)}}}\eqno(4.19)$$
in the dynamical region, III. Transformations (4.18) and (4.19) generalize the corresponding 
transformation for the Schwarzschild black hole given in ref. [14]. A proper time interval in the
dynamical region is seen to agree with the proper time interval of the asymptotic observer 
and with his Minkowski coordinate time. Its conjugate momentum, or the proper energy, will 
therefore be the energy contained in this region, between the inner and outer horizons. Now, 
the exterior and interior of the space-time admit a time-like Killing vector, $\xi^\mu$, so 
it is natural to think in terms of Komar's definition${}^{13}$ of the energy 
$$E~~ =~~ -~ {1 \over {8\pi}} \int_S *d\xi,\eqno(4.20)$$
where $S$ is a spatial two sphere. For an exterior which is vacuum, for example 
the Schwarzschild black hole, the integral is independent of the surface, $S$, and the 
definition requires only that $\xi^\mu$ is a time like Killing vector. If the surface is chosen 
to lie at infinity, Komar's definition may be applied to all asymptotically flat space-times
to obtain the total energy. In regions I, II, IV and V, for a two sphere $S$ with curvature 
radius $R$, equation (4.20) may be thought of as the total energy, $E$, of the system 
{\it interior} to $S$. The energy contained between the horizons would be the difference 
between the energies interior to the two horizons, {\it i.e.,} its energy interior to 
the outer horizon minus its energy interior to the inner horizon. Therefore, we define 
the energy $E$ in the dynamical region as 
$$E~~ =~~ -~ {1 \over {8\pi}} \int_{S_+} *d\xi~ +~ {1 \over {8\pi}} \int_{S_-} *d\xi,
\eqno(4.21)$$
where $S_-$ and $S_+$ refer to the inner and outer horizons respectively with radii $R_\mp = 
M \mp \sqrt{M^2-Q^2}$. By its definition, $E$ is a conserved quantity and a true invariant. It 
gives $E=\sqrt{M^2 - Q^2} = (R_+ -R_-)/2$ and reduces appropriately to $E=M$ in the limit as 
$Q\rightarrow 0$.
\vskip 0.1in
\centerline{\epsfysize=1.5in \epsfbox{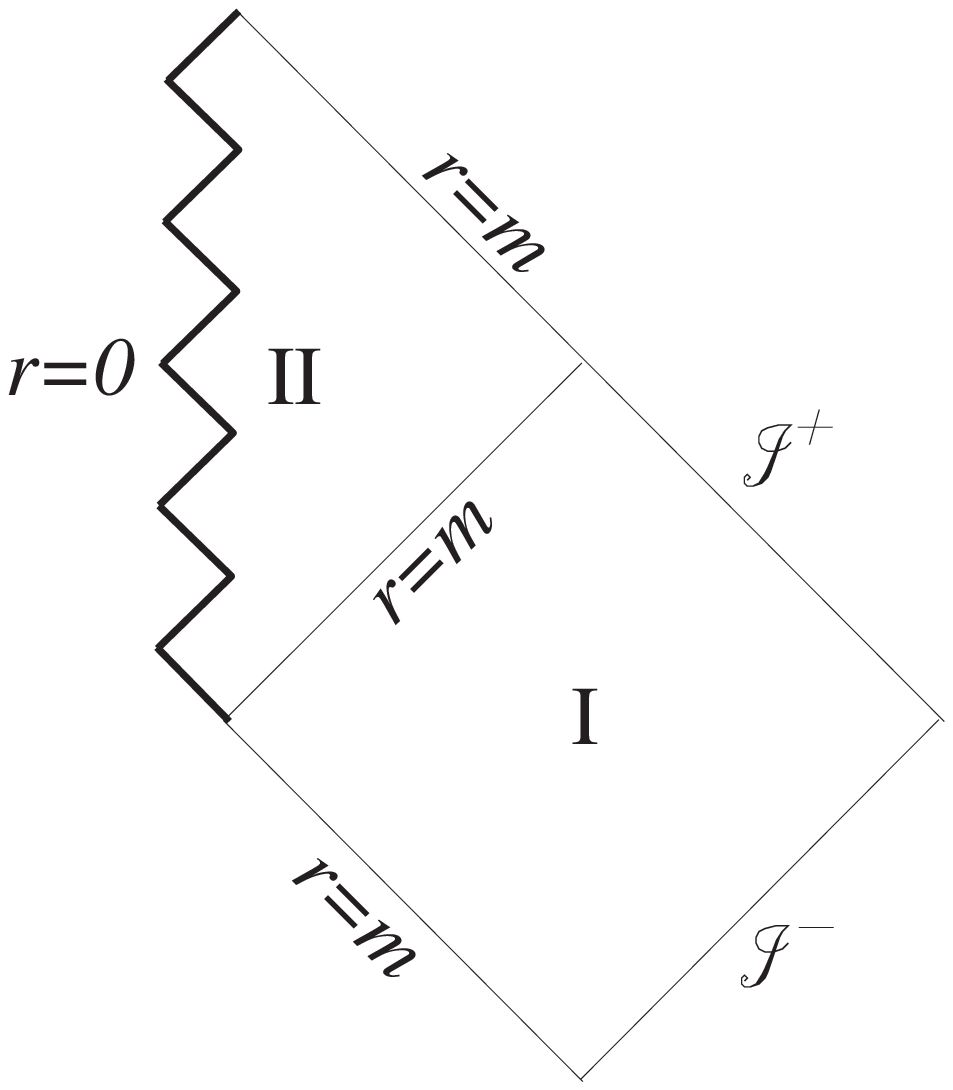}}
\vskip 0.1in
\centerline{\Tenpoint {\bf Fig. 3:} The extremal Reissner-Nordstr\"om geometry}
%\vskip 0.1in
\noindent Furthermore, region III disappears for the extremal black hole (see figure 3), 
as does the proper energy associated with it. 

Applying the definition of the energy, $E$, appropriate to this region we find the quantization 
rule
$$M\sqrt{M^2 - Q^2}~~ =~~ n M_p^2, \eqno(4.22)$$
where we have introduced the Planck mass, $M_p$. We have also thus recovered the Bekenstein 
mass spectrum for the Schwarzschild black hole ($Q \rightarrow 0$), which was examined  
separately in refs. [8,9]. Moreover, if ${\cal A}_\pm$ refer to the areas of the outer and 
inner horizons respectively, the quantization condition reads
$${\cal A}_+~ -~ {\cal A}_-~~ =~~ 16 \pi n l_p^2, \eqno(4.23)$$
{\it i.e.,} it is the {\it difference} between the outer and inner horizon areas that 
is quantized in integer units.

We must of course ensure that the wave-functionals in this region obey the supermomentum 
constraint (4.6). This is easily verified. Applying (4.6) on the wave-functional shows that
$\tau'(r) = 0 = R'(r)$, assuming that $E\neq 0 \neq a(M,Q)$. Both these conditions can be 
easily met in the dynamical region. Here there is no contradiction with the positivity of 
$L^2$ because $F<0$ and we see that the phase in (4.12) disappears leaving only the 
direct product state in (4.11). The solutions can be classified as even parity and odd 
parity states, just as in the case of the Schwarzschild black hole,
$$\eqalign{\Psi^{(+)}[\tau,R_*]~~ &=~~ \prod_r {1 \over {\sqrt{\pi M}}} e^{-iE\tau(r)} 
\cos [ER_*(r)]~~~~~ EM~ =~ (2n+1)~ ,\cr \Psi^{(-)}[\tau,R_*]~~ &=~~ \prod_r {1 \over 
{\sqrt{\pi M}}} e^{-iE\tau(r)} \sin [ER_*(r)]~~~~~ EM~ =~ 2n~ ,\cr}\eqno(4.24)$$
where we take $n \in {\rm{\bf N}} \cup \{0\}$ in keeping with the positive energy
requirement and $E=\sqrt{M^2-Q^2}$. Then if we think of the extremal 
Reissner-Nordstr\"om black hole as the limiting case of the non-extremal black hole, we 
see that it corresponds to $n=0$ with a vanishing wave functional. As a consequence, the 
entropy of the extremal black hole vanishes identically. It would 
appear that the extremal black hole either has no dynamics at all or that it cannot 
be understood as the limit $Q\rightarrow M$ of the non-extremal black hole and must be 
treated independently. It would be of interest, for example, keeping the exterior fixed, 
to consider different interior solutions that leave the horizon non-singular and contain 
modes which propagate entirely within the horizon. These modes will contribute to 
the entropy of an extremal black hole, giving a non-zero result. This is the approach 
of ref. [15], but how this program can be implemented for the non-extremal case and 
within the context of the canonical theory is not clear to us at present. 
\vskip 0.2in

\noindent{\bf V. The Entropy.}

In the previous section we have seen that the states of the black hole reside only in 
the dynamical region, $R_- < R < R_+$, between the inner and outer horizons of the black 
hole and that the Wheeler-DeWitt equation is decoupled so that the wave-functional is 
expressible as a direct product state 
$$|\Psi\rangle~~ =~~ \prod_r |\Psi_r\rangle\eqno(5.1)$$
where each component of the direct product over the label coordinate is either an 
even parity state or  an odd parity state in (4.23). We may imagine that a lattice is placed 
on the spatial hypersurface so that the classically continuous variable $r$ is a discrete
label. The Wheeler-DeWitt wave functional then represents a collection of, say, $N$ (assumed 
to be finite) decoupled oscillators each determined by the same Schroedinger equation and obeying 
the same boundary conditions. The black hole entropy is a consequence of the fact that a 
knowledge of its total  mass and charge is not equivalent to a knowledge of the number of ways 
in which this  mass and charge is distributed between the components. Each particular 
distribution corresponds to a definite microstate of the black hole. To compute the entropy 
we must enumerate these microstates.

It is therefore convenient to reformulate the problem by recognizing that the wave 
equation in the dynamical region at each label coordinate is derivable from the action
$$S_r~~ =~~ -{1 \over 2} \int_{-{{\pi M} \over 2}}^{{\pi M} \over 2} d^2 X \sqrt{-\gamma}
\gamma^{ab} \partial_a \Psi_r^\dagger \partial_b \Psi_r,\eqno(5.2)$$
where $X \in (\tau,R_*)$, and that the total action has the form
$$S~~ =~~ \sum_{r=1}^N S_r~~ =~~ -{1 \over 2} \sum_{r=1}^N \int_{-{{\pi M} \over 2}}^{{\pi M} 
\over 2} d^2 X \sqrt{-\gamma}\gamma^{ab} \partial_a \Psi_r^\dagger \partial_b \Psi_r.
\eqno(5.3)$$
The boundary conditions are that each $\Psi_r$ vanishes at the outer and inner horizons.
Performing a mode expansion of $\Psi_r$ and combining both parities, we express the contribution
of any one of the lattice sites to the total energy of the system in terms of pairs of
creation and annihilation operators, $(\alpha^\dagger_n, \alpha_n)$ and $(\beta^\dagger_n,
\beta_n)$, 
$$\eqalign{[\alpha_n,\alpha^\dagger_m]~~ &=~~ n \delta_{nm}\cr [\beta_n,\beta^\dagger_m]~~ 
&=~~ n \delta_{nm}\cr}\eqno(5.4)$$
as follows
$${\hat{\cal H}}_r~~ =~~ {{M_p^2} \over {M}} \sum_{\nr} (\anr^\dagger \anr~ +~ \bnr^\dagger
\bnr).\eqno(5.5)$$
The total energy is the sum over contributions from each of the lattice sites, {\it i.e.,}
$${\hat{\cal H}}~~ =~~ \sum_r {\hat{\cal H}}_r~~  =~~ {{M_p^2} \over {M}} \sum_{r=1}^N 
\sum_{\nr} (\anr^\dagger \anr~ +~ \bnr^\dagger \bnr)\eqno(5.6)$$
and this must be the energy of the dynamical region. We consequently obtain a dispersion 
relation of the form
$$\eqalign{E~~ &=~~  {{M_p^2} \over M} \sum_{r=1}^N \sum_{\nr,\lr} (\nr N_\nr~ +~ \lr K_\lr)\cr 
&=~~ {{M_p^2}\over M} \sum_{r=1}^N \nu_r~~ =~~ {{M_p^2}\over M}{\cal N},~~~~~ \nu_r, {\cal N}~ 
\in~ {\bf N} \cup \{0\},\cr}\eqno(5.7)$$
where $\nr$, $\lr$ and $N_\nr$, $N_\lr$ refer respectively to the level number and the 
occupation number at level number $\nr$($\lr$) corresponding to the oscillator at lattice 
site $r$. Equation (5.7) is yet another way of seeing that it is the product, $EM$, that 
is quantized in integer units.

A crude estimate of the number of states may be given as follows. Let $\rho(\nu_r)$ be the 
density of levels describing each site $r$. It will naturally account for the angular 
variables which have been integrated over in the canonical theory and have so far played 
no role. Assuming that the lattice sites are distinguishable, for a generic level density 
and large $\nu_r$ the density of states can be written as${}^{9}$
$$\Omega(E,M,N)~~ =~~ \prod_{r=1}^N \int_{\nu_o}^\infty d\nu_r \rho(\nu_r) \delta 
\left(EM - \sum_{s=1}^N \nu_s \right), \eqno(5.8)$$
where $\nu_o$ represents a lower limit on the energy attributable to each oscillator, which 
we take to be zero. The $\delta-$function restricts the limits of the $\nu_r$ integrals in the product. 
Following Carlitz${}^{16}$ and Frautschi${}^{17}$ we estimate the $r^{\rm th}$ integral by
$$\int_0^{\lambda_r} d\nu_r \rho(\nu_r),\eqno(5.9)$$
provided that the upper limits, $\lambda_r$, are subject to the condition 
$$\sum_{r=1}^N \lambda_r~~ =~~ EM.\eqno(5.10)$$
Then the maximum contribution to $\Omega(E,M,N)$ is obtained when all of the $\lambda_r$ 
are of the order $EM/N$. This provides an estimate for the integrals in (5.9). Quite 
generally
$$\Omega(E,M,N)~~ =~~ a^N f^N(\xi),\eqno(5.11)$$
where $\xi = b E M/N$ and $a$ and $b$ are constants which can be determined from the 
density of states. The precise value of these constants is irrelevant because, as 
we shall see, the functional dependence of the entropy on the mass and charge of the 
black hole is a consequence only of the dispersion relation, $EM=\sum_{r=1}^N 
\nu_r$, in (5.7). So far $N$ has been introduced by hand as the number of 
lattice points on a typical hypersurface. We determine its value by maximizing 
the number of states with respect to it. Notice that
$${{\partial} \over {\partial N}} \ln \Omega(E,M,N)~~ =~~ \ln a~ +~ \ln f(\xi)~ -~ 
\xi {\partial \over {\partial \xi}} \ln f(\xi)~~ =~~ 0\eqno(5.12)$$
is a homogeneous equation in $\xi$ whose solution will yield a value, say $\xi=\alpha^{-1}$. 
Assuming that it is real, this value determines the number of lattice points as 
$$N_{max}~~ =~~ \alpha b EM\eqno(5.13)$$
and thus the entropy as
$$S(M,Q)~~ =~~ \ln \Omega(E,M,N_{max})~~ =~~ \gamma M\sqrt{M^2 - Q^2},\eqno(5.14)$$
where $\gamma$ is determined in terms of $\alpha$, $a$ and $b$, by inserting (5.13) 
in (5.11),
$$\gamma~~ =~~ \alpha b \ln \left[a f(\alpha^{-1})\right].\eqno(5.15)$$
The entropy is the difference between the outer horizon area and the inner horizon area,
$$S~~ =~~ {{\gamma} \over {4\pi}}\left({{{\cal A}_+ - {\cal A}_-}\over 4}\right) 
\eqno(5.16)$$
and is quantized in integer units. Our result does not coincide with Bekenstein's 
proposal, $S \approx {\cal A}_+$, except if $Q\rightarrow 0$.
\vskip 0.2in

\noindent{\bf VI. Discussion.}

We have generalized our earlier study of the energy spectrum and the statistical entropy 
of the Schwarzschild black hole to the charged, Reissner-Nordstr\"om black hole. The 
canonical reduction differs only slightly from Kucha\v r's treatment of the Schwarzschild 
black hole and the quantization program is seen to lead to precisely the same wave-functional 
as was obtained in refs. [8,9] for the Schwarzschild black hole, when it is taken to vanish 
in the interior, static region. Our fundamental result is that if $E$ is the energy 
associated with the dynamical region, the quantization condition and entropy are given by
$$\eqalign{EM~~ &=~~ {\cal N} M_p^2,~~~~~ {\cal N}~ \in {\bf N} \cup \{0\},\cr S~~ &=~~ 
\gamma E M.\cr}\eqno(6.1)$$
where $\gamma$ is a constant that can be determined. The key distinction between the charged 
and uncharged black hole is in what we define to be the ``energy'' associated with the dynamical 
region. In the absence of a canonical choice and motivated by physical considerations, we have 
made what we consider to be a reasonable proposition: take it to be the energy contained between 
the inner and the outer horizons of the Reissner-Nordstr\"om solution and calculated as the 
difference between the Komar integrals evaluated on the horizons in the exterior and in the 
interior, where time-like Killing vectors exist. This then leads to the conclusion that it 
is the {\it difference} in the horizon areas that describes the entropy and that is quantized 
in integer units. The result does not agree with Bekenstein's hypothesis, but it does 
confirm the existence of an ``area quantization'' law.

It follows that the temperature of the charged black hole, as measured by an observer at 
infinity, does not agree with its semi-classical Hawking temperature. Indeed it is lower 
than the Hawking temperature due to a contribution from the inner horizon. One finds,
$$T^{-1}~~ =~~ \left({{\partial S} \over {\partial M}}\right)_Q~~ =~~ {{\gamma} \over 
2}{{R_+^2 + R_-^2} \over {\sqrt{M^2 - Q^2}}}.\eqno(6.2)$$
The first term on the right hand side is the inverse Hawking temperature and is attributable 
to the outer horizon. Likewise, the second term is an inverse ``temperature'' that is 
attributable to the inner horizon. This inner ``temperature'' is negative. If we call 
these temperatures, respectively, $T_+$ and $T_-$, the temperature of the charged black 
hole can be written as
$${1 \over T}~~ =~~ {1 \over {T_+}}~ -~ {1 \over {T_-}},\eqno(6.3)$$
where $T^{-1}_\pm = (\partial S_\pm/\partial M)_Q$ and $S_\pm = \gamma {\cal A}_\pm/16\pi$.

The canonical quantization described in this article leads to the amusing picture of a 
black hole as a microcanonical ensemble of oscillators, similar to hadrons whose statistical 
mechanics was studied many years ago by Carlitz${}^{16}$ and Frautschi${}^{17}$. The oscillators 
are imagined to be located rigidly on a lattice with $N$ sites placed on a spatial hypersurface, 
and the optimal number of sites is determined by maximizing the density of states. It turns out 
that this number is proportional to the difference in horizon areas, which leads to the expression 
we derived for the entropy. $N$ is the number of dynamical degrees of freedom of the system 
and this statistical result encourages the following interpretation. Each horizon area 
statistically yields the total number of degrees of freedom within it. An outside observer 
sees the total number of degrees of system within $R_+$ minus the total number of degrees 
of freedom within $R_-$, which difference is the number of degrees of freedom 
added dynamically.

What physical processes are responsible for the ``lost'' degrees of freedom? The temperature 
and the entropy of the black hole are eventually measured by an external, distant observer by 
placing a ``thermometer'' in the form of a test quantum field in the black hole background. 
Only those modes of the quantum field that do not enter the outer horizon 
are accessible to the asymptotic observer. A transition from a higher energy state to a lower 
energy state by the emission of a single quantum of radiation may occur on any oscillator, 
including one whose degree of freedom is attributable to the inner horizon. However, the 
quantum field in the exterior would not be sensitive to an emission from such an ``inner'' 
oscillator so such a transition would elicit no response from the ``thermometer''. 
As far as the asymptotic observer is concerned, therefore, these degrees of freedom must 
be subtracted from the total. The semi-classical theory does not account for this loss, hence 
the expression (5.16) for the entropy is lower than the semi-classical analysis would predict 
and the temperature associated with the black hole in the canonical theory lower than the 
Hawking temperature.

An equivalent way of viewing this is to observe that the semi-classical analysis does not 
involve the back reaction to any radiation emitted exterior to the black hole. Electrically 
neutral radiation, for example, would diminish the mass of the black hole but leave its charge 
unchanged. Hence $R_+$ would decrease but $R_-$ would increase. Semi-classically, the asymptotic 
observer would be unaware of this increase. He would also be unaware of a similar effect 
induced by the possible emission of radiation from the interior of the black hole, which 
radiation would either be absorbed by the singularity or pass on to another branch of the 
universe. Hence the measure of entropy and temperature given by the asymptotic observer 
appears to need adjustment or interpretation.

While it may be argued, as we have, that the semi-classical analysis is not sensitive to the 
conditions in the interior of the hole and is therefore not reliable when multiple horizons are 
involved, it is also possible that the semi-classical result is correct and that the boundary 
conditions we have imposed are too restrictive. We have treated the interior as the analytical 
extension of the exterior, but it has been suggested${}^{15}$, for example, within the string 
theory framework and for extreme black holes, that the internal region must be replaced by 
solutions which contain modes propagating entirely within it and matched smoothly to the exterior.  
At the time of writing, we do not know how such a program would be implemented for non-extremal 
black holes within the context of the canonical theory.
\vskip 0.2in

\noindent{\bf Acknowledgements.}

\noindent We acknowledge the partial support of the {\it Centro de Astrof\'\i sica} and of FCT, 
Portugal, under contract number SAPIENS/32694/99. L.W. acknowledges the hospitality of the 
Universidade do Algarve, Faro, Portugal. L.W. was supported in part by the Department of Energy 
under contract Number DOE-FG02-84ER40153.
\vfill\eject

\noindent{\bf References.}

\item{1.}J. D. Bekenstein, Ph. D. Thesis, Princeton University, April 1972; {\it ibid}, 
Lett. Nuovo Cimento {\bf 4} (1972) 737.

\item{2.}J. D. Bekenstein, Phys. Rev. {\bf D7} (1973) 2333; Phys. Rev. {\bf D9} (1974) 
3292. 

\item{3.}J. D. Bekenstein, Lett. Nuovo Cimento {\bf 11} (1974) 467.

\item{4.}S. W. Hawking, Commun. Math. Phys. {\bf 43} (1975) 199. 

\item{5.}J. D. Bekenstein, Phys. Lett. {\bf B360} (1995) 7; Phys. Rev. Lett. {\bf 70} (1993) 
3680.

\item{6.}J. D. Bekenstein, ``Black Holes: Classical Properties, Thermodynamics and Heuristic 
Quantization'', in the IX Brazilian School on Cosmology and Gravitation, Rio de Janeiro 1998, 
gr-qc/9808028; {\it ibid}, ``Quantum Black Holes as Atoms'', in the VIII Marcel Grossman 
Meeting on General Relativity, Jerusalem, June 1997, gr-qc/9710076; J. D. Bekenstein and V. 
Mukhanov in {\it Sixth Moscow Quantum Gravity Seminar}, ed. V. Berezin, V. Rubakov and 
D. Semikoz (World Publishing, Singapore, 1997); V. Mukhanov, in {\it Complexity, Entropy 
and the Physics of Information: SFI Studies in the Sciences of Complexity}, Vol III, ed. W. 
H. Zurek (Addison-Wesley, New York, 1990).

\item{7.}D. Christodoulou and R. Ruffini, Phys. Rev. {\bf D4} (1971) 3552.

\item{8.}Cenalo Vaz and Louis Witten, Phys. Rev. {\bf D60} (1999) 024009.

\item{9.}Cenalo Vaz, Phys. Rev. {\bf D61} (2000) 064017.

\item{10.}K. V. Kucha\v r, Phys. Rev. {\bf D50} (1994) 3961.

\item{11.}K. V. Kucha\v r and Charles G. Torre, Phys. Rev. {\bf D43} (1991) 419.

\item{12.}J. D. Brown and K. V. Kucha\v r, Phys. Rev. {\bf D51} (1995) 5600.

\item{13.}A. Komar, Phys. Rev {\bf 113} (1959) 934.

\item{14.}L.D. Landau and E.M. Lifshitz, {\it The Classical Theory of Fields}, 
Butterworth-Heinemann (1997).

\item{15.}G. Horowitz and D. Marolf, Phys. Rev. {\bf D55} (1997) 3654.

\item{16.}R. D. Carlitz, Phys. Rev. {\bf D5} (1972) 3231.

\item{17.}S. Frautschi, Phys. Rev. {\bf D3}, (1971) 2821.
\bye